\documentclass[11pt]{article}
\newcommand{\Comment}[1]{{}}
\usepackage[textwidth = 430 pt, textheight = 630 pt]{geometry}
\usepackage{amssymb,euscript,amsmath,amsfonts,filecontents}
\usepackage[dvipsnames]{xcolor}
\usepackage[utf8]{inputenc}
\title{ }

\usepackage{color}
\definecolor{MyDarkBlue}{rgb}{0.15,0.15,0.45}
\usepackage[linktocpage=true]{hyperref}
\hypersetup{
colorlinks=true,
citecolor=MyDarkBlue,
linkcolor=MyDarkBlue,
urlcolor=MyDarkBlue,
pdfauthor={ N. Lambert and T. Orchard},
pdftitle={Null Reductions of the M5 Brane},
pdfsubject={hep-th}}

\usepackage[numbers,sort&compress]{natbib}
\usepackage{hypernat}

\numberwithin{equation}{section}
 
\begin{document}

\renewcommand{\thefootnote}{\fnsymbol{footnote}}
 
 \centerline{\LARGE \bf {\sc Null Reductions of the M5-Brane}}
 
 \vspace{1cm}
 
   \centerline{
   {\large {\bf   {\sc Neil Lambert}}\footnote{E-mail address: \href{neil.lambert@kcl.ac.uk}{\tt neil.lambert@kcl.ac.uk}  }  \,   {\sc    and Tristan Orchard}}\footnote{E-mail address: \href{mailto:tristan.orchard@kcl.ac.uk}{\tt tristan.orchard@kcl.ac.uk}} }

\vskip 1cm
\centerline{\it Department of Mathematics}
\centerline{{\it King's College London }} 
\centerline{{\it  WC2R 2LS, UK}} 

\vspace{1.0truecm}

 
\thispagestyle{empty}

\centerline{\sc Abstract}
\vspace{0.4truecm}
\begin{center}
\begin{minipage}[c]{360pt}{
We perform a general reduction of an M5-brane on a spacetime that admits a null Killing vector, including couplings to background 4-form fluxes and possible twisting of the normal bundle. We give the non-abelian extension of this action and present its supersymmetry transformations.   The  result is a class of supersymmetric non-Lorentzian gauge theories in 4+1 dimensions, which depend on the geometry of the six-dimensional spacetime. These can be used for DLCQ constructions of M5-branes reduced on various manifolds.     \noindent}
 
\end{minipage}
\end{center}

\renewcommand{\thefootnote}{\arabic{footnote}}
\setcounter{footnote}{0}

 \newpage 

\section{Introduction}

The M5-brane is an interesting and important object in M-Theory for a variety of reasons. Its dynamics are described by a six-dimensional field theory with $(2,0)$ supersymmetry. For multiple M5-branes this is an interacting, strongly coupled superconformal field theory. However, we currently lack a satisfactory understanding of this theory.
Nevertheless, a particularly fruitful application of M5-branes involves compactifying them on a manifold to obtain lower dimensional field theories. In this way, many novel field theories have been identified as well as relations/dualities between them.  

Recently, we have studied null reductions of the M5-brane (a related abelian construction already appeared in \cite{Bandos:2008fr} as well as in newer work \cite{Townsend:2019ils}).  In the simplest cases, this leads to the construction of novel non-abelian field theories in (4+1)-dimensions with 24 (conformal) supersymmetries \cite{Lambert:2018lgt,Lambert:2019jwi}.  Due to the fact that one has fixed a particular null direction in the six-dimensional theory, the Lorentz group has been reduced from $SO(1,5)$ to $SO(4)$. However, they still admit a large bosonic spacetime symmetry, including a Lifshitz scaling, coming from the six-dimensional conformal group \cite{Lambert:2019fne}.  In this paper we extend this discussion to general null reductions of the M5-brane on a curved manifold. 

Non-Lorentzian theories with Lifshitz scaling have received a great deal of attention, primarily  from the perspective of their AdS dual geometry (for a review see \cite{Taylor:2015glc}). While some supersymmetric Lifshitz theories have been explicitly constructed (for example see \cite{Xue:2010ih,Chapman:2015wha}) these often involve higher derivative terms, as is common in condensed matter systems.  
In contrast, the field theories we obtain do not have higher derivatives but involve Lagrange multiplier constraints that reduce the dynamics to motion on a moduli space of anti-self-dual gauge fields \cite{Lambert:2011gb,Mouland:2019zjr}, in line with the DLCQ description of the M5-brane \cite{Aharony:1997an,Aharony:1997pm}. Other classes of theories without Lorentz invariance but related to String/M-Theory have recently received attention in works such as \cite{Harmark:2018cdl,Bergshoeff:2019pij,Harmark:2019zkn,Bergshoeff:2020xhv,deBoer:2020xlc,Bergshoeff:2020baa}.

 These more general null reductions should provide DLCQ-type descriptions of the field theories obtained by reducing M5-branes on other manifolds such as the Gaiotto theories \cite{Gaiotto:2009we}. 
Since there is no six-dimensional action based on non-abelian fields, the standard construction is to reduce the abelian theory and then find a suitable non-abelian extension that is compatible with supersymmetry. For example, this was performed in  \cite{Linander_2012} for the case of a general spacelike circle fibration. This was then followed by \cite{cordova2017five}, who generalised this construction to include additional non-dynamical supergravity background fields. In this paper we will apply these constructions to the case of a null reduction. 
We will not consider the full background supergravity fields that were discussed in \cite{cordova2017five}, however we will extend our results to backgrounds coming from fluxes in M-theory and a non-trivial connection on the normal bundle.

Although conceptually similar, reduction on a null direction is technically distinct from a spatial reduction and involves some interesting features. In particular in a spatial reduction on a coordinate $x^5$ the self-duality constraint on  the closed three-form $H$ is solved by eliminating  $H_{MNP}$ in terms of $F_{MN}\sim H_{MN 5}$  for $M,N,P\ne 5$. One then introduces a gauge potential for $F$ in order to formulate a Lagrangian. In contrast in a null reduction along $x^+$ 
 $H$ leads to self-dual and anti-self-dual  two-forms $G_{ij}\sim H_{ij-}$ and $F_{ij}\sim H_{ij+}$ in the remaining spatial directions $i,j=1,2,3,4$. To construct the action we  impose the self-duality of $G$ and introduce a gauge potential for $F$ off-shell. The equations of motion then imply the anti-self-duality of $F$ along with a suitable closure condition on $G$. 
 
This paper is organised as follows. 
In section two we perform the general reduction of the abelian $(2,0)$ theory equations on a general spacetime with a null isometry. While the $(2,0)$ theory is based on a tensor multiplet, upon reduction we obtain vector fields. We then generalise the resulting action to a supersymmetric non-abelian gauge theory in section three. 
  In section four we examine some special cases of the general reduction, and in section five we include couplings to background flux terms. Section six contains our conclusions and comments. Our conventions are summarised in the appendix, along with some formulae for the geometry. 

\section{The Abelian Dimensional Reduction}

 In this section we will reduce the equations of motion and supersymmetry variations of the abelian $(2,0)$ tensor multiplet on a six-dimensional manifold with metric $\hat g_{MN}$ which admits a null Killing direction $\hat{k}^M$. We will use hats to denote six-dimensional geometrical objects throughout.
 
\subsection{The Background}

Consider a fixed curved background, {\it i.e.} there is no back-reaction on the metric from the matter fields.
We will further only consider six-dimensional Lorentzian manifolds which admit a null killing vector field
\begin{align}
	\hat k  =  \frac{\partial}{\partial x^{+}} \ .
\end{align}
 In coordinates adapted to this isometry, $(x^{+}, x^{-}, x^{i} )$ $i \in \{1, \dots,4\}$ it can be shown that the metric takes the general form (see also \cite{Julia_1995})
\begin{align}\label{gdef}
    \hat{g}_{M N} = 
        \begin{pmatrix}
            0 & -1 & -u_j  \\
            -1 & -2\sigma &  -v_j - 2\sigma \, u_j \\
         -u_i  & -v_i -2\sigma \, u_i & g_{i j} -2 u_{(i}v_{j)} - 2\sigma \, u_i u_j
        \end{pmatrix}\ .
\end{align}
Here $g_{i j}$ is a Euclidean signature metric of a four-dimensional submanifold of the full six-dimensional spacetime. All components of $\hat g_{MN}$ are allowed to depend on $x^-$ and $x^i$. The metric component $g_{+-}=-1$ has been fixed using a suitable choice of the coordinate $x^-$. This somewhat contrived choice of metric  was chosen as it leads to the simpler inverse metric
\begin{align}
    \hat{g}^{M N} = 
        \begin{pmatrix}
            |v|^2 + 2 \sigma & \underline{u}\cdot \underline{v} -1 & -v^j \\
            \underline{u}\cdot \underline{v} -1 & |u|^2 & - u^j \\
            -v^i & -u^i & g^{ \, i j}
        \end{pmatrix}\ .
\end{align}
It is important to note that this geometry is distinct from that invoked in \cite{Bilal:1999ff}, in which a spacelike circle is infinitely  Lorentz boosted. Even if limits are examined carefully in that paper, one finds as the boost parameter goes to zero the length of the Killing vector is always positive. In contrast, our Killing vector has length zero, as would be expected from a null reduction. 

For the time being we do not consider any other background fields other than the metric; in section \ref{sec:flux} off-brane fluxes are added.
\subsection{Tensor Multiplet}

The six-dimensional abelian $\mathcal{N} = (2,0)$ tensor multiplet   contains a self-dual   3-form,
\begin{align}
      H = \hat{\star} H, 
\end{align}
along with five scalar fields, $X^I$, and a symplectic Majorana-Weyl spinor $\psi$. These fields transform in the trivial, fundamental, and spinor representations of the $R$-symmetry group $SO(5)$ (or equivalently $USp(4)$) respectively. 

The supersymmetry transformations
\begin{align}
\label{eq:susytrans}
    \begin{split}
        &\delta X^I = i \bar{\epsilon} \hat \Gamma^I \psi \\
        &\delta H_{MNP} = 3i\partial_{[M}( \bar{\epsilon} \hat\Gamma_{NP]} \psi) \\
        &\delta \psi = \hat D_M X^I  \hat\Gamma^M \hat\Gamma^I \epsilon + \frac{1}{2 \cdot 3!} H_{MNP}\hat\Gamma^{MNP} \epsilon\ ,
    \end{split}
\end{align}
close up to the equations of motion: 
\begin{align}
 H = \hat{\star} H\ ,\qquad 	\hat{d}  H = 0\ ,\qquad   \hat D^M\hat D_M  X^I=0\ ,\qquad \hat\Gamma^M \hat D_M \psi=0\ .
\end{align}
 Here the supersymmetry parameter $\epsilon$ has opposite chirality under $\Gamma_{012345}$ to $\psi$, we make the choice $\Gamma_{012345}\epsilon = \epsilon$ and $\Gamma_{012345} \psi = - \psi$.


\subsection{Reducing $H = \hat{\star} H$}

Let us first define the  (4+1)-dimensional fields:
\begin{align}
    F_{ij} = H_{ij+}, \qquad F_{i-} = H_{i-+}, \qquad G_{ij} = H_{ij-} \ .
\end{align}
In a trivial geometry these three fields are the independent components of the six-dimensional 3-form $H$, and $F$ and $G$ satisfy simple (anti-)self-duality constraints. Our task here is to see the implications of the six-dimensional self-duality condition for a general background. 

In what follows we use the geometrical quantities associated to the four-dimensional manifold with metric $g_{ij}$. In particular we define the fields $F^{ij}$, $G^{ij}$ and $F_{-}^{i}$ to have their indices raised by $g^{\,ij}$. We also take 
\begin{align}
	\varepsilon_{+-ijkl}=\varepsilon_{ijkl}\ ,
\end{align}
with $\varepsilon_{1234}=1$. Along with the metric $g_{ij}$, this allows  us to define a four-dimensional Hodge star operator $\star$.    
 To proceed it is convenient to work with forms, we define the one forms $v = v_i dx^i$, $u = u_i dx^i$ and $F_{-} = F_{i-} dx^i$, as well as the two forms $F = \frac{1}{2} F_{ij} dx^i \wedge dx^j$ {\it etc.}. We also define the 3-form $H = \frac{1}{3!}H_{ijk}dx^i\wedge dx^j \wedge dx^k$. \\
\\
Written in forms the self-duality condition on $F_-$ is
\begin{align}
    \label{eq:F-form}
    F_{-} = \star(v\wedge u \wedge F_{-}) + \star(v \wedge F) + \star(u \wedge G  ) - \star H\ .
\end{align}
Applying $\star$ allows us to solve for $H$
\begin{align}
    \label{eq:Hform}
    H = \star F_{-} + v\wedge u \wedge F_{-} + v \wedge F + u \wedge G \ .
\end{align}
Eliminating $H$ from the other relations   we create two equations that depend only on $F_{-}, F, G$ along with the background fields $\sigma,u,v$ and $g $. In particular we find
\begin{align}
    \begin{split}
        F = - \star \! F + F_{-} \wedge u + \star (F_{-} \wedge u)
    \end{split} \nonumber\\
    \begin{split}
        {G} = \star  {G} - 2 \sigma \star \! F - F_{-} \wedge v + \star (F_{-} \wedge v) + 2\sigma \star(F_{-} \wedge u) \ . 
    \end{split}
\end{align}
Defining 
\begin{align}
\mathcal{F} &= F - F_{-} \wedge u \nonumber\\
\mathcal{G} &= G - \sigma F - F_-\wedge (v+\sigma u) \ ,	
\end{align}
 these expressions simplify further to
\begin{align}
    \begin{split}
        \mathcal{F} = - \star \! \mathcal{F}
    \end{split} \nonumber \\
    \begin{split}
       {\cal  G} = \star {\cal G}\ .
    \end{split}
\end{align}


\subsection{Decomposing $\hat{d}H = 0$}

The exterior derivative is metric independent, so the results will hold for all backgrounds. In components
\begin{align}
    \partial_{[M}H_{NPQ]} = 0\ .
\end{align}
Our construction has a $x^{+}$ isometry, so all fields are independent of $x^+$. This gives an expression for each of the combinations of indices $+\! -\! ij, \ +ijk, \ -ijk, \ ijkl$
\begin{align}
    \partial_{[+}H_{-ij]} = 0 \quad \implies& 
    \quad   \partial_{-}F  + dF_- = 0
    \nonumber  \\
    \partial_{[+} H_{ijk]} = 0 \quad \implies& 
   \quad dF = 0 
    \nonumber\\
    \partial_{[-}H_{ijk]} = 0 \quad \implies& 
    \quad d {G} = \partial_{-} H 
    \nonumber\\
    \partial_{[i} H_{jkl]} = 0 \quad \implies& \quad dH = 0.
\end{align}
Where we have written the 4 dimensional exterior derivative as  $d$. The first and second expressions can be combined to give a simple five-dimensional Bianchi identity  
\begin{align}
\label{eq:dF=0}
  \quad d_{(5)}F_{(5)} = 0 \ ,\qquad F_{(5)} = F + F_-\wedge dx^-\ .
\end{align}
Implying that locally there exists $(A_-, A_i)$ such that 
\begin{align}\label{eq:Adef}
    F_{ij} = \partial_{i} A_{j} - \partial_{j} A_{i}\ , \qquad F_{i -} = \partial_{i} A_{-} - \partial_{-}A_{i}\ .
\end{align}
The equations for $\mathcal {G}$ and $\mathcal{F}$ become
\begin{align}
\begin{split}
\label{eq:dH}
    d(\mathcal {G} + \sigma\mathcal{F} - F_{-}\wedge v ) = \partial_{-} ( \star F_{-} + v\wedge u \wedge F_{-} + v \wedge F + u \wedge (\mathcal {G} + \sigma \mathcal{F} - F_{-} \wedge v))
\end{split} \nonumber \\
\begin{split}
     d(\star F_{-} + v\wedge u \wedge F_{-} + v \wedge F + u \wedge (\mathcal {G} + \sigma \mathcal{F} - F_{-} \wedge v)) = 0\ .
\end{split}
\end{align}
Using the duality properties of $\mathcal {F}$ and $\mathcal {G}$ we can rewrite these equations in component form as
\begin{align}
    \begin{split}
    \label{eq:delta A}
        D_j \mathcal{G}^{i j} + D_j \big( \sigma \star \mathcal{F}^{ij} \big) - D_j \big( \star ( F_{-} \wedge v) ^{ij} \big) + D_{-} F^{i}_{\ -} - D_{-} \big( \star F^{ij}v_j \big) - D_{-} \big( \sigma \star \mathcal{F}^{ij} u_j \big) \\ - D_{-} \big( \mathcal {G}^{ij}u_j \big) = 0 \nonumber
    \end{split}
\end{align}
\begin{align}
    -D_i F^{i}_{\ -} + D_i \big( \star \! F^{ij}v_j \big) + D_i \big( \mathcal {G}^{ij} u_j \big) + D_i \big( \sigma \star \! \mathcal{F}^{ij} u_j \big) = 0\ ,
\end{align}
respectively.
\subsection{An Action}

Lastly we wish to construct an action that reproduces these equations of motion, along with those of the scalars and fermions. In the latter cases a six-dimensional action already exists which can be trivially reduced to find an appropriate five-dimensional action.

Somewhat remarkably the  equations for $F_-, F$ and $\mathcal {G}$ can be derived from a Lagrangian density on a four-dimensional manifold with Euclidean signature, whose fields also depend on the `time' coordinate $x^-$. To this end we assume that $F_-$ and $F$ arise from a potential $(A_-,A_i)$ as in (\ref{eq:Adef}). However, we do not impose a potential for $\mathcal {G}$ but rather impose $\mathcal{G}=\star\mathcal{G}$ \footnote{Note that this is a legitimate imposition, as self-dual tensors are an irreducible representation of the Lorentz group in even dimension}. Some trial and error shows that the equations of motion (\ref{eq:delta A}) then arise from the lagrangian
\begin{align}
    \mathcal{L}_H = \frac{1}{2} \star \! F_{-} \wedge F_{-} - \frac{1}{2} \sigma \star \! \mathcal{F} \wedge \mathcal{F} + \frac{1}{2}   \mathcal{F} \wedge {\mathcal G}  - \frac{1}{2} F \wedge F_{-} \wedge v \ .
\end{align}
Where
\begin{align}
F_{ij} &= \partial_i A_j  -\partial_j A_i\nonumber\\
F_{i-} & = \partial_i A_--\partial_-A_i\nonumber\\ 
\mathcal{F}_{ij} & = F_{ij} +  2u_{[i}F_{j]-}\nonumber\\
 \mathcal{G}_{ij} &= \frac12 \sqrt{g}\varepsilon_{ijkl}\mathcal{G}^{kl}\ ,
\end{align}
and  the $k,l$ indices are raised with respect to $g^{ij}$.  Variation with respect to  $\mathcal{ G}$ immediately gives the anti-self-dual condition ${\cal F}=-\star{\cal F}$. On the other hand, varying $A_{i}$ and $A_{-}$ give (\ref{eq:delta A}) respectively. 

Inclusion of the scalars and fermions is easier, as there is a Lagrangian formulation for the free conformal case in any dimension;
\begin{align}
\mathcal{L}_{matter} = -\sqrt{-\hat g}\left(\frac12\hat g^{MN} \partial_M X^I\partial_N X^I +\frac{1}{8}\frac{d-2}{d-1} \hat R X^IX^I- \frac i2 \bar\psi\hat\Gamma^M \hat D_M\psi\right)	\ .
\end{align}
Performing the reduction by assuming $\partial_+=0$, and inserting $d=6$, we find
\begin{align}
    \mathcal{L}_{matter}  =  &- \sqrt{g} \left(\frac12\partial_i X^I \partial^i X^I -  \frac12 |u|^2 \partial_{-}X^I \partial_{-} X^I +   u^i \partial_i X^I \partial_{-}X^I - \frac{1}{10} \hat{R} X^I X^I\right.  \nonumber\\
     &\qquad \quad -\left. \frac{i}{2}  \bar{\psi} \hat \Gamma^{-}\hat{D}_{-} \psi - \frac{i}{2}  \bar{\psi}\hat \Gamma^{i}\hat{D}_{i} \psi -\frac{1}{2}i\bar\psi\hat M\psi \right)\ ,
\end{align}
where
\begin{align}
 \hat M &=	    \frac14\hat \Gamma^{+} \hat\omega_{+ MN}\hat \Gamma^{MN}\nonumber\\
 & = \frac14\partial_-u_i \hat \Gamma^{+} \hat\Gamma^{-i} + \frac14 \partial_iu_j\hat \Gamma^{+} \hat\Gamma^{ij}   \ .
\end{align}
Note that we keep the fermionic terms and $\hat R$  in their six-dimensional form. In principle these can be computed from the expression (\ref{eq:viel}), (\ref{SpinC}) and (\ref{GammaRel}) found in the appendix.  However, expanding everything out in full detail for a general background leads to rather unwieldy expressions. Rather, we will provide more explicit  expressions in various special cases below.
 
It is helpful to introduce 
\begin{align} 
\nabla_i = \partial_i -u_i \partial_-\ ,	
\end{align}
This  derivative  generally has torsion;
\begin{align}
    \nabla_{[i} \nabla_{j]} X^I = - 2 \nabla_{[i}u_{j]} \partial_{-}X^I\ .
\end{align}
One also finds that 
\begin{align}
\label{eq:bianchi}
    \nabla_{[i}\mathcal{F}_{jk]} = F_{- [i}(\partial_{j} u_{k]} - u_{j} \partial_{-} u_{k]}) \ .
    \end{align}
Putting all these together we can write the full abelian action as
\begin{align}
\begin{split}\label{AAction}
    S = \frac{1}{{g^2_{\text{YM}}}}\int d x^{-} d^4 x \sqrt{g}\Big\{ \frac{1}{2} F_{i -} F^{i}_{-} - \frac{1}{4} \sigma \mathcal{F}_{ij}\mathcal{F}^{ij} + \frac{1}{2} {\cal G}_{ij}\mathcal{F}^{ij} - \frac{1}{2\sqrt{g}} \varepsilon^{ijkl}  F_{i-} v_j F_{kl}\\  - \frac{1}{2} {\nabla}_{i} X^I  {\nabla}^i X^I  
     - \frac{1}{10}\hat{R} X^I X^I  + \frac{1}{2}i \bar{\psi} \Gamma^{-}\hat{D}_{-} \psi + \frac{1}{2}i \bar{\psi} \Gamma^{i}\hat{\nabla}_{i} \psi + \frac{1}{2}i\bar\psi\hat M\psi      \Big\}\ .
    \end{split}
\end{align}

\section{Supersymmetry and Non-Abelian Generalization}

Next we want to show that the action (\ref{AAction}) is supersymmetric. To this end 
we assume there there exists a solution to the conformal Killing spinor equation
\begin{align}
\hat D_M\epsilon = \hat\Gamma_M\eta	\ ,
\end{align}
with $\partial_+\epsilon=0$. In particular this implies
\begin{align}\label{econ}
    \hat{D}_{+} \epsilon = \frac{1}{4} \hat{\omega}_{+MN}\hat{\Gamma}^{MN} \epsilon = \hat{\Gamma}_{+} \eta\ ,
\end{align} 
which is a further condition that we must impose on the geometry.
As it stands the action (\ref{AAction}) is not invariant under the transformations that follow directly from (\ref{eq:susytrans}). One problem is that the variation $\delta \mathcal{G}_{ij}$ obtained from (\ref{eq:susytrans}) is not self-dual off-shell. Thus, we must adjust the algebra in a way that ensures $\delta \mathcal{G}_{ij}$ is self-dual. 

A deeper issue  is that  although we impose the isometry $\partial_+\psi=0$, this does not imply that $\hat{D}_{+}\psi=0$. For the bosonic fields, this distinction does not cause a problem as both $X^I$ and $H_{MNP}$ do not couple to the spacetime connection (due to the fact that $H_{MNP}$ is anti-symmetric). But for $\psi$ this leads to the  Scherk-Schwarz-like mass term 
$\frac{i}{2}\bar\psi\hat M\psi$ in (\ref{AAction}).

On-shell this is also not a problem as  $\delta H_{MNP}$ in (\ref{eq:susytrans}) contains terms involving $\hat D_+\psi $ which lead to the closure of the algebra and invariance of the equations of motion. However, we find that the $\bar \psi\hat M\psi $ term  can only be made supersymmetric in general by modifying the variation of $F_{-i}$ and $F_{ij}$ in a way that means they are no longer closed. This in turn implies that a suitable expression for the supersymmetry variation of the gauge field cannot be defined. Since the existence of such a gauge field was crucial for the construction of the action, having no definable variation is not tenable.

Alternatively, one might question why we start with the supersymmetry algebra (\ref{eq:susytrans}) and not simply
\begin{align} 
    \begin{split}
        &\delta X^I = i \bar{\epsilon} \hat \Gamma^I \psi \\
        &\delta B_{MN} = 2i   \bar{\epsilon} \hat\Gamma_{MN} \psi \\
        &\delta \psi = \hat D_M X^I \hat\Gamma^M\hat\Gamma^I  \epsilon + \frac{1}{2 \cdot 2!} \partial_MB_{NP}\hat\Gamma^{MNP} \epsilon\ ,
    \end{split}
\end{align}
identify $H = dB$ and impose $H=\hat\star H$ as an equation of motion. However, in this case one finds that $G_{ij} = 2\partial_{[i} B_{j]-} + \partial_-B_{ij}$ and hence imposing an off-shell self-duality constraint on $\mathcal{G}_{ij}$ and $\delta \mathcal{G}_{ij}$ becomes non-trivial. 

Thus, to obtain a supersymmetric action after reduction on $x^+$ we find ourselves in a balancing act of finding off-shell expressions for $\delta A_-,\delta A_i$ and $\delta \mathcal{G}_{ij}=\star \delta \mathcal{G}_{ij}$ when $\hat D_+\psi\ne0$.

\subsection{Correcting $\delta\mathcal {G}$}

The  next problem is that $\delta\mathcal {G}$ is not self-dual off-shell, but to write the action we require that $\mathcal {G}$ is self-dual. A short calculation shows that
\begin{align}
    \delta \mathcal{G}_{ij} - \star \delta \mathcal{G}_{ij} = i \bar{\epsilon} \Gamma_{-} \Gamma_{ij} E(\psi)\ ,
\end{align}
where $E(\psi)$ denotes the fermion equation of motion. Therefore,  we simply shift $\delta \mathcal{G}_{ij} \longrightarrow \delta'\mathcal{G}_{ij}  = \delta \mathcal{G}_{ij} - \frac{1}{2} i \bar{\epsilon} \Gamma_{-} \Gamma_{ij} E(\psi) $, resulting in
\begin{align}
\delta' \mathcal{G}_{ij} - \star \delta' \mathcal{G}_{ij} = \delta \mathcal{G}_{ij} - \frac{1}{2} i \bar{\epsilon} \Gamma_{-} \Gamma_{ij} E(\psi)  - \star \big( \delta \mathcal{G}_{ij} - \frac{1}{2} i \bar{\epsilon} \Gamma_{-} \Gamma_{ij} E(\psi) \big ) = 0\ ,
\end{align}
relabelling $\delta' \mathcal{G}_{ij}$ to $\delta \mathcal{G}_{ij}$ gives us a self-dual $\delta\mathcal{G}$. \\
\\
Next the $\sigma\mathcal{F}_{ij} \mathcal{F}^{ij}$ term, not present in the flat theory, must be accounted for in the supersymmetry transformations. This leads to a variation of the form
\begin{align}\label{newdL}
\delta {\cal L} = -\frac12 \sigma \mathcal{F}^{ij}\delta \mathcal{F}_{ij}  \ .
\end{align}
We must use properties of $\mathcal{F}_{ij}$ to shift $\delta \mathcal{G}_{ij}$ in such a way to cancel the effects of this new term, whilst ensuring $\delta \mathcal{G}_{ij}$ remains self-dual. \\
\\
It is useful to note that a fermionic term of definite duality, {\it e.g.} $\bar{\epsilon} \Gamma_{+} \Gamma_{ij} \psi$, can be used to build other terms of either the same or opposite duality (see Appendix A for for origin of these dualities). For instance, inserting an additional $\Gamma_{k}$ will result in either a term of the same duality; $\bar{\epsilon} \Gamma_{+} \Gamma_{k} \Gamma_{ij} \psi$, or opposite duality; $\bar{\epsilon} \Gamma_{+}  \Gamma_{ij} \Gamma_{k} \psi$. . With this in mind we choose the shift
\begin{align}
\label{eq:Gshift}
    \delta \mathcal{G}_{ij} \longrightarrow \delta \mathcal{G}_{ij} + i \bar{\epsilon} \sigma \Gamma_{+} \Gamma_{ij} \Gamma_{k} \hat{\nabla}^{k} \psi \ ,
\end{align}
which is self-dual by construction. A simple Gamma matrix manipulation shows the overall change to $\delta \mathcal{L}$ is
\begin{align}
    \delta \mathcal{L} \longrightarrow \delta \mathcal{L} + \frac{1}{2} \mathcal{F}^{ij}( i\bar{\epsilon} \sigma \Gamma_{+}\Gamma_{ijk} \hat{\nabla}^{k} \psi + 2i \bar{\epsilon} \sigma \Gamma_{+[i} \hat{\nabla}_{j]} \psi ) 
\end{align}
The last term cancels (\ref{newdL}) so we require the second term to vanish for supersymmetry. This is trivially true if $\sigma = 0$. Looking at the form of (\ref{eq:bianchi}) for $\mathcal{F}$ under $\nabla$, the first term is a total derivative if $du - u \wedge \partial_{-} u = 0$ and $\partial_i\sigma=0$. These three possibilities: $\sigma=0$ or $du - u \wedge \partial_{-} u = 0$ and $\partial_i\sigma=0$ or $\Gamma_+\epsilon=0$, arise in the two natural cases studied below.  Note that $\delta \mathcal{G}_{ij}$ also has a term proportional to $\eta$, to account for terms arising from integration by parts. \\
\\
Our corrected supersymmetry transformations read
\begin{align}
    \begin{split}
    &\delta X^I = i \bar{\epsilon}\Gamma^{I} \psi \\
    &\delta A_i \; = - i \bar{\epsilon}(\Gamma_{+-}u_i + \Gamma_{+i}) \psi \\
    &\delta A_{-} = - i \bar{\epsilon} \Gamma_{+-} \psi \\
    &\delta \mathcal{G}_{ij}  =     - \frac{1}{2} i \bar{\epsilon} \Gamma_{+} \Gamma_{-} \Gamma_{ij} \hat{D}_{-} \psi - \frac{1}{2 }i \bar{\epsilon} \Gamma_{-}\Gamma^{k} \Gamma_{ij} \hat{\nabla}_{k} \psi  + i \bar{\epsilon} \sigma \Gamma_{+} \Gamma_{ij} \Gamma^{k} \hat{\nabla}_{k} \psi - 3 i \bar{\eta}\Gamma_{-} \Gamma_{ij} \psi \\
    & \delta {\psi} \ \ \: \! = -F_{i-}  \Gamma^{+-i}{\epsilon}+ \frac{1}{4} \mathcal{F}_{ij}   \Gamma^{+ij}{\epsilon} + \frac{1}{4}\mathcal{G}_{ij} \Gamma^{-ij}{\epsilon} + \Gamma^{-} \Gamma^I  {D}_{-}X^I {\epsilon}   
    + \Gamma^i \Gamma^I \hat{\nabla}_i X^I{\epsilon} - 4 X^I \Gamma^I{\eta}  \ .
    \end{split}
\end{align}
Again we have kept many of the fermionic terms in their six-dimensional form for notational simplicity.  With these supersymmetry transformations we find that the action (\ref{AAction}) is invariant up to terms arising from the $\bar\psi\hat M\psi$ term, and terms from the shift to $\delta G$. In other words we find $\delta S=0$ if 
\begin{align}\label{Mconstraint}
	\delta\bar \psi \hat M\psi =0
	\end{align}
	and
	\begin{align}\label{OtherConstraints}
	 \sigma &= 0  \nonumber\\
	  \text{or} \quad du - u \wedge \partial_{-} u &= 0\ {\rm and}\   \partial_i\sigma=0\\ 
	  \text{or} \quad du - u \wedge \partial_{-} u &= 0\ {\rm and}\  \Gamma_+\epsilon=0\ .\nonumber
\end{align}
The implications of these constraints are explored in section \ref{sec:M}, and as we will see there is a remarkable amount of redundancy between them.

\subsection{Non-Abelian Theory}

Our next task is to find a non-abelian extension of the abelian action found above which remains supersymmetric. After some trial and error we find that, assuming (\ref{Mconstraint}) and (\ref{OtherConstraints}) holds,   non-abelian extension is 
\begin{align}
\begin{split}
    S = \frac{1}{{g^2_{\text{YM}}}}\text{tr} \int d x^{-} d^4 x \sqrt{g}\Big\{& \frac{1}{2} F_{i -} F^{i}_{-} - \frac{1}{4} \sigma \mathcal{F}_{ij}\mathcal{F}^{ij} + \frac{1}{2} \mathcal{G}_{ij}\mathcal{F}^{ij} - \frac{1}{2\sqrt{g}} \varepsilon^{ijkl}  F_{i-} v_j F_{kl} \\
   & - \frac{1}{2}  {\nabla}_{i} X^I  {\nabla}^i X^I  - \frac{1}{10}\hat{R} X^I X^I+\frac{i}{2}\bar\psi\hat M\psi \\
    &+ \frac{1}{2}i \bar{\psi} \Gamma^{-}\hat{D}_{-} \psi + \frac{1}{2}i \bar{\psi} \Gamma^{i}\hat{\nabla}_{i} \psi + \frac{1}{2} \bar{\psi} \Gamma_{+} \Gamma^I \big [ X^I, \psi \big] \Big\}\ ,
    \end{split}
\end{align}
where all the fields now live in the adjoint of some gauge group. The supersymmetry transformations are 
\begin{align}
    \begin{split}
    &\delta X^I = i \bar{\epsilon}\Gamma^{I} \psi \\
    &\delta A_i \; = - i \bar{\epsilon}(\Gamma_{+-}u_i + \Gamma_{+i}) \psi \\
    &\delta A_{-} = - i \bar{\epsilon} \Gamma_{+-} \psi \\
    &\delta \mathcal{G}_{ij}   =      - \frac{1}{2} i \bar{\epsilon} \Gamma_{+} \Gamma_{-} \Gamma_{ij} \hat{D}_{-} \psi - \frac{1}{2 }i \bar{\epsilon} \Gamma_{-}\Gamma^{k} \Gamma_{ij} \hat{\nabla}_{k} \psi  + i \bar{\epsilon} \sigma \Gamma_{+} \Gamma_{ij} \Gamma^{k} \hat{\nabla}_{k} \psi \\  
    & \qquad \quad -\frac{1}{2}i \bar{\epsilon} \Gamma_+\Gamma_- \Gamma_{ij} \Gamma^{I} \big[ X^I, \psi \big] - 3 i \bar{\eta}\Gamma_{-} \Gamma_{ij} \psi \\
     & \delta {\psi} \ \ \: \! = -F_{i-}  \Gamma^{+-i}{\epsilon}+ \frac{1}{4} \mathcal{F}_{ij}   \Gamma^{+ij}{\epsilon} + \frac{1}{4}\mathcal{G}_{ij} \Gamma^{-ij}{\epsilon} + \Gamma^{-} \Gamma^I  {D}_{-}X^I {\epsilon}   
    + \Gamma^i \Gamma^I \hat{\nabla}_i X^I{\epsilon} \\ 
      & \qquad \quad + \frac{i}{2}   \Gamma_{+} \Gamma^{IJ} \big [X^I, X^J \big]{\epsilon} - 4 X^I \Gamma^I{\eta}  \ ,
    \end{split}
\end{align}
where again we have left $\hat R$ and the fermion derivatives in their six-dimensional form. 

\subsection{Twisting}
We can also introduce an non-zero connection on the R-symmetry of the form
\begin{align}
\hat {\cal D}_M X^I &= \hat \partial_M X^I + \hat {A}_M(X^I)\nonumber\\
\hat {\cal D}_M\psi	& = \hat D_M\psi + \frac14 \hat \Omega_M^{IJ}\Gamma^{IJ}\psi\ ,
\end{align}
 and similarly for $\hat {\cal D}_M\epsilon$. This will allow us to introduce a twisting of the normal bundle.   Here $\hat A_M$ acts on $X^I$ in a representation of some subgroup $Q$ of $SO(5)$ and $\hat \Omega^{IJ}_M$ provides a spinor embedding of $Q$ into $Spin(5)$. 
 
Since this modification  only affects the dynamics through derivatives of the scalars and fermions, we can see its effect by modifying the matter part of the six-dimensional action to
\begin{align}
S_{matter} = 	 {\rm tr}\int d^6x \sqrt{-\hat g}\left( -\frac12 {\cal D}_M X^I {\cal D}^M X^I + \frac{i}{2}\bar\psi\hat\Gamma^M{\cal D}_M\psi  -\frac{1}{10}\hat R X^IX^I - \frac{1}{2}  T^{IJ}X^IX^J\right)\ ,
\end{align} 
where $T^{IJ}$ is an invariant tensor of $Q$. 
This modification leads to
\begin{align}
\delta S_{matter}  = {\rm tr}\int d^6x \sqrt{-\hat  g}\Big(\frac{i}{10}\bar\psi \Gamma^{IJK}\Gamma^{MN}\epsilon \hat {\cal R}_{MN}{}^{JK} X^I &+ \frac{3i}{10}\bar\psi \Gamma^I\Gamma^{MN}\epsilon \hat {\cal R}_{MN}{}^{IJ} X^J\nonumber\\ & - iT^{IJ}\bar\psi \Gamma^IX^J\Big)\ ,
\end{align}
where $\hat {\cal R}_{MN}{}^{IJ}$ is the curvature of $\hat \Omega^{IJ}_M$. Thus, to obtain a supersymmetric reduction we must ensure $\hat {\cal D}_M\epsilon=\hat\Gamma_M\eta$, $\partial_+\epsilon=0$ and arrange for suitable choices of curvature and $ T^{IJ}$ so that the terms in $\delta S_{matter}$ cancel. Indeed, the usual role  of twisting is to allow for solutions to $\hat {\cal D}_M\epsilon=0$ on manifolds with non-vanishing curvature.  For example, in the case of a Riemann surface along $x^3,x^4$ with normal directions $X^6,X^{7}$ the first term vanishes and we can arrange to cancel the last two by taking
\begin{align}
T^{67}	 = \mp\frac{3}{5}\hat {\cal R}_{34}{}^{67}\ ,
\end{align}
and projecting on to spinors with $\hat \Gamma_{34}\hat\Gamma^{67}\epsilon =\pm \epsilon$, where the sign is chosen to correspond to solutions of $\hat {\cal D}_M\epsilon=0$.

\section{Examples}

In the previous section we constructed the non-abelian extension of the reduced M5-brane equations and their supersymmetry transformations. We left the fermion terms in a six-dimensional form as the complete expression in full generality is quite complicated and unenlightening. In this section we will evaluate some general classes of examples explicitly. 

\subsection{Obstruction from  $\hat M$}
\label{sec:M}

In order to obtain a supersymmetric reduction we require in addition that (\ref{Mconstraint}), {\it i.e.} $\delta\bar\psi\hat M\psi=0$, is satisfied.  
In addition the condition (\ref{econ})  ensures that
\begin{align}\label{e+conditions}
\frac14\partial_-u_i\Gamma^{-i}\epsilon_- + \frac14(\partial_iu_j-u_i\partial_-u_j)\Gamma^{ij}\epsilon_+ & =  \Gamma_+\eta\nonumber\\
\partial_iu_j\Gamma^{ij}\epsilon_- & = 0\ .
\end{align}
 We do not propose to give the general solutions to these conditions which place various restrictions on both $\epsilon$ and the background fields $\sigma,u,v$. For example, if $du$ is not anti-self-dual then the second equation implies that $\epsilon_-=0$.

Since there are no mass terms for the scalars (beyond the usual conformal coupling to the curvature)  a physically well-motivated  class of background  that ensures (\ref{Mconstraint}) are those for which there is also no mass term for the fermions:
\begin{align}
	\bar\psi\hat M\psi =0\ .
\end{align}
This leads to the following conditions on the background fields
\begin{align}
du - u\wedge \partial_-u &  =-\star(du - u\wedge \partial_-u)\nonumber\\
\partial_-u & = -2i_v(du - u\wedge \partial_-u)	\nonumber\\
 \sigma(du - u\wedge \partial_-u) &=  \frac12 (1-\star)(v\wedge \partial_-u)\ .
\end{align}
With $i_v(\cdot)$ denoting contraction with $v$. We also had a further constraint that had two choices, arising from our requirement to only shift $\delta G$ by self dual terms. One recalls
\begin{align}
\sigma = 0 \quad \text{or} \quad du - u \wedge \partial_{-} u = 0 \ .
\end{align}
There are two natural solutions to these constraints:\footnote{Note that  in case 2 we could consider the the weaker conditions $du=0$ and $\partial_-u=0$. But this implies $u=df$ in which case we can set $u=0$ by a diffeomorphism $x^-\to x^-+f$.}
\begin{align}
{\rm case }\ &1: u\ne 0\ ,\partial_{-}u =0 \quad\implies \quad  v = \sigma = 0\ ,\ du= -\star du
\nonumber\\
{\rm case }\ &2:  u = 0\ ,\   v ,\sigma \ne  0\ .\end{align}
Therefore, from (\ref{e+conditions}) we find
\begin{align}
{\rm case\ 1}&: \epsilon_-\ne 0\quad \eta = -  \frac18\partial_iu_j\Gamma_-\Gamma^{ij}\epsilon_+ \nonumber\\ 	
{\rm case\ 2}&: \eta=0\ .
\end{align}

In what follows we will only focus on these two cases so that we can be as explicit as possible. We emphasize that other solutions to the constraints (\ref{econ}) and (\ref{Mconstraint}) might also be possible. 


\subsection{Case 1: $\partial_{-}u = v = \sigma = 0\quad du= -\star du$}
Here the action is
\begin{align}
\begin{split}
    S = \frac{1}{{g^2_{\text{YM}}}}\int d x^{-} d^4 x \sqrt{g}\Big\{ \frac{1}{2} F_{i -} F^{i}{}_{-} + \frac{1}{2}  \mathcal{G}_{ij}\mathcal{F}^{ij} - \frac{1}{2} \nabla_{i} X^I \nabla^i X^I - \frac{1}{10}\hat{R} X^I X^I \\ +\frac{1}{2}i\bar{\psi}\Gamma^{-}D_{-} \psi  + \frac{1}{2}i \bar{\psi} \Gamma^{i} \nabla_i \psi - \frac{1}{4} e^{\underline{i}}_{\ [i} \partial_{-}e_{j]\underline{i}}  \bar{\psi}(\Gamma^{-} - u_k \Gamma^k) \Gamma^{ij} \psi \\ + \frac{1}{2} \bar{\psi} \Gamma_{+} \Gamma^{I} \big [ X^I, \psi \big ]  \Big\},
    \end{split}
\end{align}
which is invariant under
\begin{align}
    \begin{split}
    &\delta X^I = i \bar{\epsilon}\Gamma^{I} \psi \\
    &\delta A_i \; = - i \bar{\epsilon}(\Gamma_{+-}u_i + \Gamma_{+i}) \psi \\
    &\delta A_{-} = - i \bar{\epsilon} \Gamma_{+-} \psi \\
    &\delta   \mathcal{G}_{ij}  =   - \frac{1}{2} i \bar{\epsilon} \Gamma_{+} \Gamma_{-} \Gamma_{ij} D_{-} \psi - \frac{1}{2 }i   \bar{\epsilon} \Gamma_{-}\Gamma^{k} \Gamma_{ij}( D_{k} - u_k D_{-}) \psi - \frac{1}{2}i \partial_{-} g_{kl} \bar{\epsilon} \Gamma^k \Gamma_{ij} \Gamma^l \psi_{-}    \\ 
    & \qquad \quad \, - \frac{1}{4}i e^{\underline{i}}_{\ [k} \partial_{-} e_{l] \underline{i}} \bar{\epsilon} \Gamma_{ij} \Gamma^{kl} \psi_{+} - \frac{1}{8}i e^{\underline{i}}_{\ [k} \partial_{-} e_{l] \underline{i}} u_p \bar{\epsilon} \Gamma^p \Gamma_{ij} \Gamma^{+} \Gamma^{kl} \psi - 3 i \bar{\eta}\Gamma_{-} \Gamma_{ij} \psi  \\
     & \qquad \quad \, -\frac{1}{2}i \bar{\epsilon} \Gamma_+\Gamma_- \Gamma_{ij} \Gamma^{I} \big[ X^I, \psi \big] \\
   & \delta {\psi} \ \ \: \! = -F_{i-}  \Gamma^{+-i}{\epsilon}+ \frac{1}{4} \mathcal{F}_{ij}   \Gamma^{+ij}{\epsilon} + \frac{1}{4}\mathcal{G}_{ij} \Gamma^{-ij}{\epsilon} + \Gamma^{-} \Gamma^I  {D}_{-}X^I {\epsilon}   
    + \Gamma^i \Gamma^I  {\nabla}_i X^I{\epsilon} \\
    & \qquad \quad \, + \frac{i}{2}    \Gamma_{+} \Gamma^{IJ} \big [X^I, X^J \big]{\epsilon} - 4 X^I \Gamma^I{\eta}  \ .
    \end{split}
\end{align}
 For brevity we have left the six-dimensional Ricci scalar unexpanded, for completeness in terms of four-dimensional objects only this is 
\begin{align}
\begin{split}
\hat{R} = R &- \frac{1}{2}g^{ij}\big( \partial_{-}^2 g_{ij} + \frac{1}{2}|u|^2 g^{kl} \partial_{-}g_{ik} \partial_{-}g_{jl} - g^{kl} \partial_{-} g_{ik} u_m \gamma^m_{\ jl}\big) \\
          &-u^i \big( \partial_j g^{jk} \partial_{-} g_{ki} + g^{jk}\partial_{-}g_{ik} \gamma^{l}_{\ kl} - g^{jk} \partial_{-} g_{kl} \gamma^{l}_{\ ji} - \partial_{-} \gamma^{j}_{\ ij} + \frac{1}{2}\partial_i (g^{jk}\partial_{-} g_{jk}) \big),
\end{split}
\end{align}
with $ \gamma^{i}_{ \ jk}$ the Christoffel symbols of the 4d metric. In the specific case of this metric being independent of $x^{-}$ this reduces to 
\begin{align}
    \hat{R} = R\ .
\end{align}

\subsection{Case 2: $u=0$}
\begin{align}
    \begin{split}
    S = \frac{1}{{g^2_{\text{YM}}}}\int d x^{-} d^4 x \sqrt{g}\Big\{ \frac{1}{2} F_{i -} F^{i}{}_{-} - \frac{1}{4} \sigma F_{ij}F^{ij} + \frac{1}{2} {\mathcal G}_{ij}F^{ij} - \frac{1}{2\sqrt{g}} \varepsilon^{ijkl}  F_{i-} v_j F_{kl} \\ - \frac{1}{2} D_{i} X^I D^i X^I 
    + \frac{1}{2} i \bar{\psi} \Gamma^{-}D_{-} \psi + \frac{1}{2} i \bar{\psi} \Gamma^{i} D_{i} \psi - \frac{1}{4}(\partial_{[i} v_{j]} + e^{\underline{i}}_{\ [i} \partial_{-}e_{j]\underline{i}}) \bar{\psi} \Gamma^{-ij} \psi \\   + \frac{1}{2} \bar{\psi} \Gamma_{+} \Gamma^{I} \big [ X^I, \psi \big ] \Big \},
    \end{split}
\end{align}
since $u$ is now zero $\mathcal{F} = F$. Note also that since $\eta = 0$, we have $\hat D_M\epsilon=0$ and hence $\hat R=0$.   This action is invariant under the following transformations
\begin{align}
    \begin{split}
    &\delta X^I = i \bar{\epsilon}\Gamma^{I} \psi \\
    &\delta A_i \; = - i \bar{\epsilon} \Gamma_{+i} \psi \\
    &\delta A_{-} = - i \bar{\epsilon} \Gamma_{+-} \psi \\
    &\delta \mathcal{G}_{ij}  =   - \frac{1}{2} i \bar{\epsilon} \Gamma_{+} \Gamma_{-} \Gamma_{ij} D_{-} \psi + \frac{1}{2}i (\partial_{-} v_k - \partial_{k} \sigma) \bar{\epsilon} \Gamma_{ij} \Gamma^{-k} \psi - \frac{1}{8} i (\partial_k v_l - e^{\underline{i}}_{\ k} \partial_{-} e_{\underline{i} l}) \bar{\epsilon} \Gamma_{ij} \Gamma^{kl} \Gamma_{+} \Gamma_{-} \psi \\  
    & \qquad \quad - \frac{1}{2 }i \bar{\epsilon} \Gamma_{-}\Gamma^{k} \Gamma_{ij} D_{k} \psi - \frac{1}{4} i (\partial_k v_l - \frac{1}{2} \partial_{-} g_{kl} ) \bar{\epsilon} \Gamma^{k}\Gamma_{ij} \Gamma^{l} \Gamma_{-} \Gamma_{+} \psi  + i \bar{\epsilon} \sigma \Gamma_{+} \Gamma_{ij} \Gamma^{k} D_{k} \psi \\
    & \qquad \quad -\frac{1}{2}i \bar{\epsilon} \Gamma_+\Gamma_- \Gamma_{ij} \Gamma^{I} \big[ X^I, \psi \big] \\
        & \delta {\psi} \ \ \: \! = -F_{i-}  \Gamma^{+-i}{\epsilon}+ \frac{1}{4} {F}_{ij}   \Gamma^{+ij}{\epsilon} + \frac{1}{4}\mathcal{G}_{ij} \Gamma^{-ij}{\epsilon} + \Gamma^{-} \Gamma^I  {D}_{-}X^I {\epsilon}   
    + \Gamma^i \Gamma^ID_i X^I{\epsilon} \\
    & \qquad \quad  + \frac{i}{2}    \Gamma_{+} \Gamma^{IJ} \big [X^I, X^J \big]{\epsilon}  \ .  
    \end{split}
\end{align}
However, we remind the reader that, due to (\ref{OtherConstraints}), the action is only invariant under $\epsilon_+\ne 0$ if $\sigma$ is independent of $x^i$.


\section{Flux Terms}
\label{sec:flux}

In \cite{cordova2017five} the reduced M5-branes action is coupled to  background supergravity fields such as  a non-zero M-theory 4-form $\hat G_{\mu\nu\rho\sigma}$.\footnote{The authors of \cite{cordova2017five} use a $USp(4)$ notation where the flux terms are denoted by $S^{mn}$ and  $T^{mn}_{ab}$ with $m,n=1,2,3,4$ and  $a,b=0,1,2,3,4,5$.}   The presence of such a flux leads to Myers-like terms in the M5-brane effective action. In addition the fluxes modify the Killing spinor condition to: 
\begin{align}\label{MKSE}
0=\hat D_\mu \epsilon  + \frac{1}{288}\left(\hat G_{\nu\lambda\rho\sigma}\hat \Gamma^{\nu\lambda\rho\sigma}{}_\mu+ 8\hat G_{\mu\nu\lambda\rho}\hat \Gamma^{\nu\lambda\rho}\right)\epsilon\ .	
\end{align}
We need to find fluxes that are compatible with the condition $\partial_+\epsilon=0$. In particular applying the condition $\partial_+\epsilon=0$ to (\ref{MKSE})  for the choice $\mu=+$ leads to a purely algebraic constraint. For simplicity we will restrict our attention here to cases where this constraint is trivial: {\it i.e.} $\hat D_+\epsilon=0$ and there is no contribution in (\ref{MKSE}) from the fluxes for $\mu=+$. Non-trivial cases arise in case 1 and require a cancellation between $\hat D_+\epsilon$ and the fluxes or twisting (and perhaps including additional restrictions on $\epsilon$). These are better addressed on a case-by-case basis rather than our general discussion. Thus, we restrict to case 2  ($u=0$), where $\hat D_+\epsilon=\partial_+\epsilon=0$. From the form of the above Killing spinor equation, it is easy to see that we need only consider constant fluxes of the form
\begin{align}
\hat G_{\mu\nu\lambda-} =  C_{\mu\nu\lambda} \ ,
\end{align}
with $\mu,\nu,\lambda\ne +,-$. In particular we find the possibilities $C_{IJK},C_{IJk}, C_{Ijk}$ and $C_{ijk}$, with all other combinations identically zero. These are expected to lead to additional terms in the M5-brane effective action of the form:
\begin{align}
	S' \sim \frac{1}{{g^2_{\text{YM}}}}{\rm tr}\int d x^- d^4 x\sqrt{g}  \Big(& C^{IJK} X^I[X^J,X^K] +    C^{IJi} X^ID_iX^J + C^{Iij} X^IF_{ij}   \nonumber \\ &    +  C^{ijk}\left(A_i\partial_j A_k - \frac{2i}{3}A_iA_jA_k\right) - \frac{1}{2}m^2_{IJ}X^IX^J +\frac{i}{2}\bar \psi m\psi \Big)\ , 
	\end{align}
where  $m$ and $m_{IJ}$ are a masses which are linear in the fluxes.
 Starting with a general ansatz, we find only the following corrections to the action can be made supersymmetric:
\begin{align}\label{Sprime}
	S'= \frac{1}{{g^2_{\text{YM}}}}{\rm tr}\int d x^- d^4 x\sqrt{g}\left(\     \frac{1}{6}C^{Iij} X^IF_{ij}+\frac{i}{144}\bar \psi \left( -\Gamma_+C^{IJK}\Gamma^{IJK}+ 3 \Gamma_+C^{Ijk}\Gamma^{I}{}\Gamma_{jk} \right)\psi\right)\ . 
	\end{align}
Along with this there are additional terms in the supersymmetry transformations: $\delta\to \delta+\delta'$ with
\begin{align}
 \delta' \psi &  =  -\frac{1}{12}   C^{JKL}\Gamma^{IJKL}\Gamma_+X^I\epsilon -\frac{1}{6}    C^{IJK}\Gamma^{JK}\Gamma_+X^I \epsilon \nonumber\\ 
&\qquad   -\frac13  C^{Ijk}\Gamma_{jk}\Gamma_+X^I\epsilon -\frac14  C^{Ijk}\Gamma_+\Gamma_{jk}\Gamma^{IJ}X^J\epsilon \nonumber\\
\delta'G_{ij} & = -\frac{7i}{144} C^{IJK} \bar\epsilon\Gamma_+\Gamma_{ij}\Gamma_- \Gamma^{IJK}\psi + \frac{i}{12} (C^I+\star C^I)_{ij}\bar\epsilon \Gamma_-\Gamma_+ \Gamma^I\psi \nonumber\\ &\qquad 
-  \frac{5i}{24}C^{Ikl}\bar\epsilon\Gamma_+\Gamma_- \Gamma^{I} \Gamma_{kl}\Gamma_{ij}\psi	
-\frac{i}{48}  C^{Ikl} \bar\epsilon\Gamma_+\Gamma_-\Gamma^{I}\Gamma_{ij}\Gamma_{kl}\psi \ ,
\end{align}
and furthermore the Killing spinor equation is modified to 
\begin{align}
\hat{D}_i\epsilon &= \frac{1}{72} C^{IJK}\Gamma^{IJK}\Gamma_+\Gamma_i\epsilon   -\frac16  C^I{}_{ik}\Gamma^{I}\Gamma^k \Gamma_+\epsilon  - \frac{1}{24} C^{Ijk}\Gamma^{I}\Gamma_{ijk} \Gamma_+\epsilon \nonumber\\
	\hat{D}_-\epsilon &=  
	\frac{1}{72} C^{IJK}\Gamma^{IJK}\Gamma_{+ -}\epsilon+\frac{1}{36} C^{IJK}\Gamma^{IJK} \epsilon    + \frac{1}{24} 	 C^{Ijk}\Gamma^I\Gamma_{jk}\Gamma_{+ -}\epsilon  +\frac{1}{12}  	 C^{Ijk}\Gamma^I\Gamma_{jk}\epsilon  \nonumber\\ 
	\hat{D}_+\epsilon &= 0\ ,
\end{align}
which is in agreement with the eleven-dimensional supergravity Killing spinor equation (\ref{MKSE}).  

 At first glance our result is somewhat surprising: we find no supersymmetric corrections possible for fluxes of the form $C^{ijk}$ or $C^{IJk}$, no Myers-type flux term for $C^{IJK}$ and no bosonic mass terms at all. One way to see this strange behaviour is to note that the null theory can be obtained from a non-Lorentzian rescaling of   familiar five-dimensional Yang-Mills theory \cite{Lambert2019rishi}. Here one makes the rescaling of space and time according to
\begin{align}
	x^i \to  \zeta^{-1/2} x^i , \qquad x^0\to \zeta^{-1} x^0 \ ,
\end{align}
and the matter fields by
\begin{align}
X^I \to \zeta X^I\ ,\qquad  \psi_+\to \zeta^{3/2}\psi_+\ ,\qquad \psi_-\to \zeta \psi_-\ ,  	
\end{align}
and then takes the limit $\zeta\to 0$, carefully removing divergent terms. One then makes the identification $x^-=x^0$ (but note that $\Gamma_{-} = (\Gamma_0-\Gamma_5)/\sqrt{2}$). 
The scaling of the supersymmetry parameter $\epsilon$ is fixed by requiring the fields scale the same way as their supersymmetry variations, this leads to \cite{Lambert2019rishi}
\begin{align}
\epsilon_+ \to \epsilon_+\ ,\qquad \epsilon_-\to \zeta^{-1/2}\epsilon_-	\ .
\end{align}
 
Let us now consider the form of $S'$ that would arise from a spacelike reduction of the M5-brane  in a non-vanishing supergravity  flux ({\it e.g.} as in \cite{cordova2017five}):
\begin{align}
	S'_{SYM} \sim \frac{1}{{g^2_{\text{YM}}}}{\rm tr}&\int   d^5 x       \sqrt{g}\Big(  C^{IJK} X^I[X^J,X^K]+  C^{IJM}X^ID_MX^J  + C^{IMN} X^IF_{MN}\nonumber\\ &+  C^{MNP}\left(A_M\partial_N A_P - \frac{2i}{3}A_MA_NA_P\right) - \frac{1}{2}m^2_{IJ}X^IX^J +\frac{i}{2}\bar\psi m \psi   \Big)\ ,
\end{align}
where again $m_{IJ}$ and $m$ are linear in the fluxes.  Examining the Killing spinor equation (\ref{MKSE}) one sees that we must scale the fluxes according to
\begin{align}
C_{\mu\nu\lambda} \to \zeta^{-1}	C_{\mu\nu\lambda}\ ,
\end{align}
otherwise we encounter divergences or the fluxes are scaled away. 
As a result, the deformed action scales as, schematically,
\begin{align}
S'_{SYM}  \sim &\frac{1}{{g^2_{\text{YM}}}}{\rm tr}\int d x^- d^4 x    \sqrt{g}\Big(  
 \zeta C^{IJK} X^I[X^J,X^K] +    C^{Iij} X^IF_{ij}\nonumber\\ & +  \zeta^{ 1/2} C^{IJi}X^ID_iX^J+  \zeta^{-1/2} C^{ijk}\left(A_i\partial_j A_k - \frac{2i}{3}A_iA_jA_k\right)\nonumber\\   
 &+ i \psi_-^T C_{\mu\nu\lambda} \Gamma^{\mu\nu\lambda}\psi_- + i\zeta \psi_+^T C_{\mu\nu\lambda} \Gamma^{\mu\nu\lambda}\psi_+    -   \zeta  C^{I\nu\lambda}C^J{}_{\nu\lambda}X^IX^J   \Big)\ .
\end{align}
Thus, in the limit $\zeta\to 0$, the only terms in $S'$ that survive are precisely those in (\ref{Sprime}). The only exception is the Chern-Simons-like term which diverges, and therefore is not consistent with taking the limit. \\

\section{Conclusions and Comments}

In this paper we performed a general reduction of the M5-brane along a null Killing direction. We then extended the result to a non-abelian theory. The result is a class of supersymmetric gauge theories in 4+1 dimensions but without Lorentz invariance. We also explored the effect of coupling of background supergravity fluxes to the M5-brane and  twistings of the normal bundle.

The results presented above include and generalise  earlier results. In particular simply setting $u = v = \sigma = 0$ and $g_{ij} = \delta_{ij}$ recovers the flat space case \cite{Lambert:2018lgt}, and setting $u_i = \frac{1}{2} \Omega_{ij}x^j$ recovers the metric and action of of \cite{Lambert:2019jwi}.

An interesting feature of this construction is how the information of $H$ is encoded in a consistent way into the Lagrangian. Our isometry creates a natural split in the field; $H_{ij+} = F_{ij}$, $H_{i-+}=F_{i-}$ and  $H_{ij-} =  {G}_{ij}$. $H$ is self-dual and closed, which is  problematic for a Lagrangian. But here we find  ${F}$ is closed but with no self-duality constraint off-shell, whereas $ { G}$ satisfies a  self-duality constraint but is not closed. On-shell the self-duality of $ {G}$ enforces anti-self-duality condition on  ${ F}$ as its equation of motion. In effect we have introduced a Lagrange multiplier, but without adding any new unphysical fields to our Lagrangian; $H$ provides its own Lagrange multiplier. It would be interesting to explore how this construction ties in with the six-dimensional lagrangian approach of \cite{Sen:2019qit,Lambert:2019diy,Andriolo:2020ykk}.
 
 In case 2 ${\mathcal G}$ imposes the constraint $F = -\star F$, and therefore the dynamics is restricted to the space of anti-self-dual gauge fields on the four-dimensional submanifold. Such field configurations are then solved for by the ADHM construction in terms of moduli. The remaining part of the action leads to one-dimensional motion on the instanton moduli space \cite{Lambert:2011gb,Lambert2019rishi}. This is in keeping with the various DLCQ proposals such as \cite{Aharony:1997an,Aharony:1997pm}. In case 1, ${\mathcal G}$ imposes the constraint ${\mathcal F} = -\star {\mathcal F}$ but here there are time-derivative terms and hence there is no simple reduction to motion on a moduli space but it would be interesting to explore the resulting constraint.
 
 The general form for the action includes an $F \wedge F_{-} \wedge v$ term which we can think of as a mixed Chern-Simons term between diffeomorphisms and gauge transformations. In particular for  case 1 this term vanishes but in 
in case 2, we have  $u=0$  and  so ${\cal F}_{ij} =F_{ij}$. In this case if we let  $v_{(5)} = v_i dx^i + \sigma dx^-$  then the metric admits a diffeomorphism $x^+\to x^+ + \omega$ which has the effect of mapping $v_{(5)}\to v_{(5)} +d_{(5)}\omega$ where $\omega$ depends on $x^i$ and $x^-$.\footnote{Curiously this diffeomorphism allows us to set $\sigma=0$ even though manifest supersymmetry of the action can be affected by the choice of $\sigma$.}  We can rewrite the terms involving $F$ as
\begin{align}
	\mathcal{L}_{F}  = \frac12  {\rm tr}(F_-\wedge \star F_-) -\frac18 \sigma  {\rm tr}\big((F-\star F)\wedge \star (F
	-\star F)\big)  + \frac12 {\rm tr} (F\wedge {\cal G}) +\mathcal{L}_{\text{cs}} \ ,
\end{align}
where   \begin{align}
\label{eq:CS2}
    \mathcal{L}_{\text{cs}} =  - \frac{1}{4} {\rm tr}(  F_{(5)}\wedge   F_{(5)}) \wedge  v_{(5)}\ ,
\end{align}
 and $  F_{(5)} = F + F_{-} \wedge dx^-$. Thus, under a diffeomorphism $v_{(5)} \to v_{(5)}+ d_{(5)}\omega$ the Lagrangian shifts by a total derivative. Alternatively we can write 
\begin{align}
     \mathcal{L}_{\text{cs}} =   \frac{1}{4}{\rm tr}\left( A_{(5)}\wedge dA_{(5)} - \frac{2i}{3} A_{(5)} \wedge A_{(5)}\wedge A_{(5)}\right) \wedge dv_{(5)}\ ,
\end{align}
in which case the gauge symmetry is only preserved up to a boundary term. 
We cannot write this term in a way which makes explicit both of these invariances simultaneously. Thus, we see that $\mathcal{L}_{\text{cs}} $  mixes a five-dimensional diffeomorphism with the $U(1)$ part of the gauge symmetry. 

We hope that the results will be of use in studying the $(2,0)$ and related theories reduced on non-trivial manifolds through DLCQ-type constructions \cite{Aharony:1997an,Aharony:1997pm}. For example, one could consider theories of class ${\cal S}$ \cite{Gaiotto:2009we} obtained by reduction of M5-branes on a Riemann surface $\Sigma$.  Our results here should allow for a systematic construction in terms of motion on the moduli space of instantons on ${\mathbb R}^2\times \Sigma$, {\it i.e.} Hitchin systems, coupled to scalars, fermions and possible additional data associated with singularities of $\Sigma$.


\section*{Acknowledgements}

We would like to thank Rishi Mouland for discussions. N.L. was support in part by STFC grant  ST/L000326/1 and would like to thank the CERN Theory Division for hospitality. T.O.  is supported by the STFC studentship ST/S505468/1.  
 
\section*{Appendix A: Conventions}

In this paper our conventions we use $\mu,\nu=0,1,2,...,10$ and consider an M5-brane with worldvolume coordinates $x^M$, $M=0,1,2,...,5$. However, we also introduce light cone coordinates 
\begin{align}
	x^{+} = \frac{1}{\sqrt{2}} (x^0 + x^5)\ ,\qquad  x^{-} =  \frac{1}{\sqrt{2}}  (x^0 - x^5)\ ,\qquad x^i\ , \ i=1,2,3,4\ .
\end{align}
We will use hats to denote six-dimensional geometrical quantities.

Fermions are dealt with by using Gamma matrices that satisfy a flat Clifford algebra in eleven dimensions (again with light cone Minkowski metric). All other Gamma matrices appearing in our work are derived from this basis as outlined below. Underlined indices refer to the tangent space.
\begin{center}
\begin{tabular}{ |c|c|c|c| } 
 \hline
 Notation & Definition & Description & Indices\\
 \hline
 & & & \\
 
 $\Gamma^{\underline{\mu}}$ & $\{ \Gamma^{\underline{\mu}}, \Gamma^{\underline{\nu}} \}  = 2 \eta^{\, \underline{\mu \nu}}$ & Matrices of Spin(1,10) &$\underline{\mu} \in \{0, \dots, 10\}$ \\ 
 & & & \\
 $\Gamma^{\underline{M}}$ & $\{ \Gamma^{\underline{M}}, \Gamma^{\underline{N}} \}  = 2 \eta^{\, \underline{MN}}$ & On the brane &$\underline{M} \in \{+,-,1, \dots, 4\}$ \\ 
 & & & \\
 $\Gamma^{I}$ & $\{ \Gamma^{I}, \Gamma^{J} \}  = 2 \delta^{\, \underline{IJ}}$ & Off the brane &$I \in \{6, \dots, 10\}$ \\ 
 & & & \\
 $\hat{\Gamma}^M$ & $\hat{e}^{M}_{\ \underline{M}} \Gamma^{\underline{M}}$ & 6d curved index Gamma matrices & $M \in \{+,-,1,\dots,4$\} \\ 
 & & & \\
 $\Gamma^{i}$ & $e^{i}_{\ \underline{i}} \Gamma^{\underline{i}}$ & 4d curved index Gamma matrices & $i \in \{1,\dots,4\}$\\
 & & & \\
 \hline
\end{tabular}
\end{center}
To avoid the confusion of whether or not $\Gamma^{\underline{+}}$ means $\Gamma^{\underline{\text{plus}}}$ or $   \Gamma^{\text{plus minus}}$, we will only use 
\begin{align}
	\Gamma^{+}  &= \frac{\Gamma^{\underline 0}+\Gamma^{\underline 5}}{\sqrt{2}}\ , \qquad \Gamma^{-}  = \frac{\Gamma^{\underline 0}-\Gamma^{\underline 5}}{\sqrt{2}}\nonumber\\
	\Gamma_{+}  &= \frac{\Gamma_{\underline 0}+\Gamma_{\underline 5}}{\sqrt{2}}\ , \qquad \Gamma_{-}  = \frac{\Gamma_{\underline 0}-\Gamma_{\underline 5}}{\sqrt{2}}\ .
\end{align}
The relations 
\begin{align}
    \begin{split}\label{GammaRel}
        &\hat{\Gamma}^+ = \Gamma^{+} - \sigma \Gamma^{-} - v_{i}  \Gamma^{i}, \qquad \hat{\Gamma}^{-} = \Gamma^{-} - u_{i}\Gamma^{i} \qquad \hat{\Gamma}^{i} = \Gamma^{i} \\
        &\hat{\Gamma}_{+} = \Gamma_{+}, \qquad \hat{\Gamma}_{-} = \sigma \Gamma_{+} + \Gamma_{-}, \qquad \hat{\Gamma}_{i} = (v_i + \sigma u_i) \Gamma_{+} + u_i \Gamma_{-} + \Gamma_i\ ,
    \end{split}
\end{align}
will be repeatedly used. \\
\\
The subscript $\pm$ on spinors labels their eigenvalue under $\Gamma_{\underline{05}}$, {\it e.g.}:
\begin{align}
\Gamma_{\underline{05}}\epsilon_\pm = \pm\epsilon_\pm	\ .
\end{align}
In addition we always have that $\Gamma_{\underline{012345}}\epsilon=\epsilon$ and $\Gamma_{\underline{012345}}\psi=-\psi$. This has the crucial consequence of giving certain spinor bilinears definite duality under the 4d Hodge star. Consider the following spinor bilinear
\begin{align}
\bar{\epsilon} \Gamma_{ij} \psi \ .
\end{align}
Since $\Gamma_{\underline{012345}}\psi=-\psi$, it follows that
\begin{align}
\Gamma_{\underline{12}}\psi = \Gamma_{\underline{34}} \Gamma_{\underline{05}} \psi \ ,
\end{align}
or in general 
\begin{align}
\Gamma_{\underline{ij}}\psi = \frac{1}{2} \varepsilon_{\underline{ijkl}}\Gamma^{\underline{kl}} \Gamma_{\underline{05}} \psi \ .
\end{align}
From this its easy to see that $\Gamma_{ij}\psi_{+}$ is self-dual, while $\Gamma_{ij}\psi_{-}$ is anti-self-dual under the four-dimensional Hodge star. Since $\epsilon$ has the opposite chirality under $\Gamma_{\underline{012345}}$, these are reversed: $\Gamma_{ij}\epsilon_{+}$ is anti-self-dual, $\Gamma_{ij}\epsilon_{-}$ is self-dual.
\section*{Appendix B: The Background}

The vielbein (and inverse) for the metric are given by $\hat{e}^{\ \underline{M}}_{M} \,  \hat \eta_{\underline{M N}} \, \hat{e}^{\underline{N}}_{\ N} = \hat{g}_{M N} $, with $\eta_{\underline{MN}}$ the light-cone Minkowski metric in six dimensions. This results in
\begin{align}
\label{eq:viel}
    \hat{e}^{\, \underline{M}}_{\ M} = 
        \begin{pmatrix}
            1 & \sigma & v_i + \sigma \, u_i \\
            0 & 1 & u_i \\
            0 & 0 & e^{\, \underline{i}}_{\ i}
        \end{pmatrix}\ ,
   \qquad 
   \hat{e}^{\, M}_{\ \underline{M}} =
        \begin{pmatrix}
            1 & -\sigma & -v_{\underline{i}} \\
            0 & 1 & -u_{\underline{i}}  \\
            0 & 0 & e^{\, i}_{\ \underline{i}}
        \end{pmatrix},
\end{align}
with $e^{\, \underline{i}}_{\ j}$ being the veilbien for the four-dimensional metric $g_{\, ij}$.
Where $u^i$ and $v^i$ are defined to have their index raised by $g_{\, ij}$, such that dot products are defined also with $g_{\, ij}$. We also note that
\begin{align}
    \hat{g} = \det(\hat{g}_{MN}) = \det(\hat{e}^{\underline M}{}_{N})^2 \det(\hat\eta_{\underline{MN}}) = -\det(g_{ij})\ .    \end{align}
\\
Adding the fermions requires knowledge of the spin connection terms, the non zero terms of which are
\begin{align}
    \begin{split}\label{SpinC}
        \hat{\omega}_{+-i}  &= \frac{1}{2}\partial_{-}u_i \\
        \hat{\omega}_{+ij}  &= \partial_{[i}u_{j]} \\
        \hat{\omega}_{-+i} &= \frac{1}{2} \partial_{-}u_i \\
        \hat{\omega}_{--i} &= -\partial_{i} \sigma + u_i \partial_{-} \sigma + 2 \sigma \partial_{-} u_i + \partial_{-} v_i\\
         \hat{\omega}_{-ij} &= \partial_{[i} (v_{j]} + 2 \sigma u_{j]}) + u_{[i} \partial_{-} v_{j]} - v_{[i} \partial_{-} u_{j]} - e^{\underline{i}}_{\ [i} \partial_{-} e_{|\underline{i}| j]}  \\
         \hat{\omega}_{i+-} &= - \frac{1}{2}\partial_{-}u_i \\
         \hat{\omega}_{i+j} &= \partial_{[i}u_{j]} \\
         \hat{\omega}_{i-j} &= \partial_{[i}( v_{j]} + 2 \sigma u_{j]}) + 2 u_{(i} \partial_{j)} \sigma + \partial_{-} (u_{(i}(v_{j)} + \sigma u_{j)})) - \frac{1}{2}\partial_{-}g_{ij} \\
         \hat{\omega}_{ijk} &=  \omega_{ijk} + \partial_{j} ( u_{(i}(v_{k)} + \sigma u_{k)})) - \partial_{k} ( u_{(i}(v_{j)} + \sigma u_{j)})) + \partial_{i}( u_{[j} v_{k]}) + 2 (v_{[j} + \sigma u _{[j}) \partial_{|i|} u_{k]}\ ,
    \end{split}
\end{align}
where $\omega_{ijk}$ is the four-dimensional spin connection for $D_i$, the Levi-Civita connection for $g_{ij}$ on our euclidean submanifold.

\bibliographystyle{JHEP}
\bibliography{ref}

\end{document}